\documentclass[a4paper]{jpconf}
\usepackage{graphicx}
\usepackage{amsmath}
\usepackage{amssymb}

\usepackage{citesort}

\usepackage[utf8]{inputenc}

\begin{document}
\title{Constraints on the hadronic spectrum from Lattice QCD}

\author{P Parotto}

\address{Department of Physics, University of Houston, Houston, TX 77204, USA}

\ead{pparotto@uh.edu}

\begin{abstract}
The spectrum of hadronic resonances continually receives updates from the Particle Data Group, which lists every state with a  status representing how established the state is. Moreover, the existence of additional states is predicted by relativistic quark models. It has been suggested that further states might need to be included in the hadronic spectrum in order to improve the agreement between the hadron resonance gas model predictions and lattice QCD data. Such an inclusion would also affect the results of many areas of heavy-ion collision physics that make use of hadronic degrees of freedom, such as hydrodynamical simulations afterburners. However, for some selected observables, the inclusion of further states worsens the agreement with the lattice results. We propose new observables, sensitive to the spectrum content divided by quantum numbers, which allow us to gauge the contribution of additional states. The comparison of Lattice QCD results and predictions from the Hadron Resonance Gas model for these observables, helps to clarify the situation and determine how many, and which new states are needed.
\end{abstract}

\section{Introduction}

The precision reached by Lattice simulations of QCD thermodynamics has allowed to study with increasing level of detail the properties of strongly interacting matter. Observables like fluctuations of conserved charges \cite{Borsanyi:2011sw,Bazavov:2012jq}, sensitive to the chemical composition of a strongly interacting system can now be calculated with sufficient precision, allowing for an in-depth analysis of the relevant degrees of freedom near the QCD transition \cite{Bellwied:2013cta,Bazavov:2013dta}. 

For temperatures below the QCD transition, in the hadronic phase, most Lattice results can be reproduced successfully by the Hadron Resonance Gas (HRG) model, whose underlying assumption is that a gas of interacting hadrons in the ground state can be well approximated by a gas on non-interacting hadrons and resonances. The original idea of Rolf Hagedorn, was that a hadronic system responds to an arbitrary increase in energy density, not through and indefinite increase in the temperature, but rather populating the system via the creation of resonances \cite{Hagedorn:1965st}. Thus, the concept of a limiting temperature for strongly interacting matter naturally arises, which now is well understood to be the limiting temperature for \textit{hadronic} matter, hence connected to the transition temperature of QCD \cite{Cabibbo:1975ig,Aoki:2006br,Bazavov:2011nk}. For this reason, it is reasonable to expect the effects of resonance proliferation to become more relevant at the occurrence of such transition.
 
The HRG model containing no interaction, provides the most advantageous setup, being the evaluation of any thermodynamic quantity a simple sum over the hadron and resonance spectrum of the theory. In this sense, the model is effectively parameter free, and depending only on the content of the spectrum of hadronic states that is included in the sum. However, the content of the hadronic spectrum, especially in the strange sector, is far from being established. The Particle Data Group (PDG) continuously updates the list of experimentally discovered resonances \cite{Patrignani:2016xqp}, and the existence of each resonance is assigned a confidence level, where **** indicate the most established, and * indicates those with the least experimental confirmation.

\begin{figure}[h]
\center
\includegraphics[width=14pc]{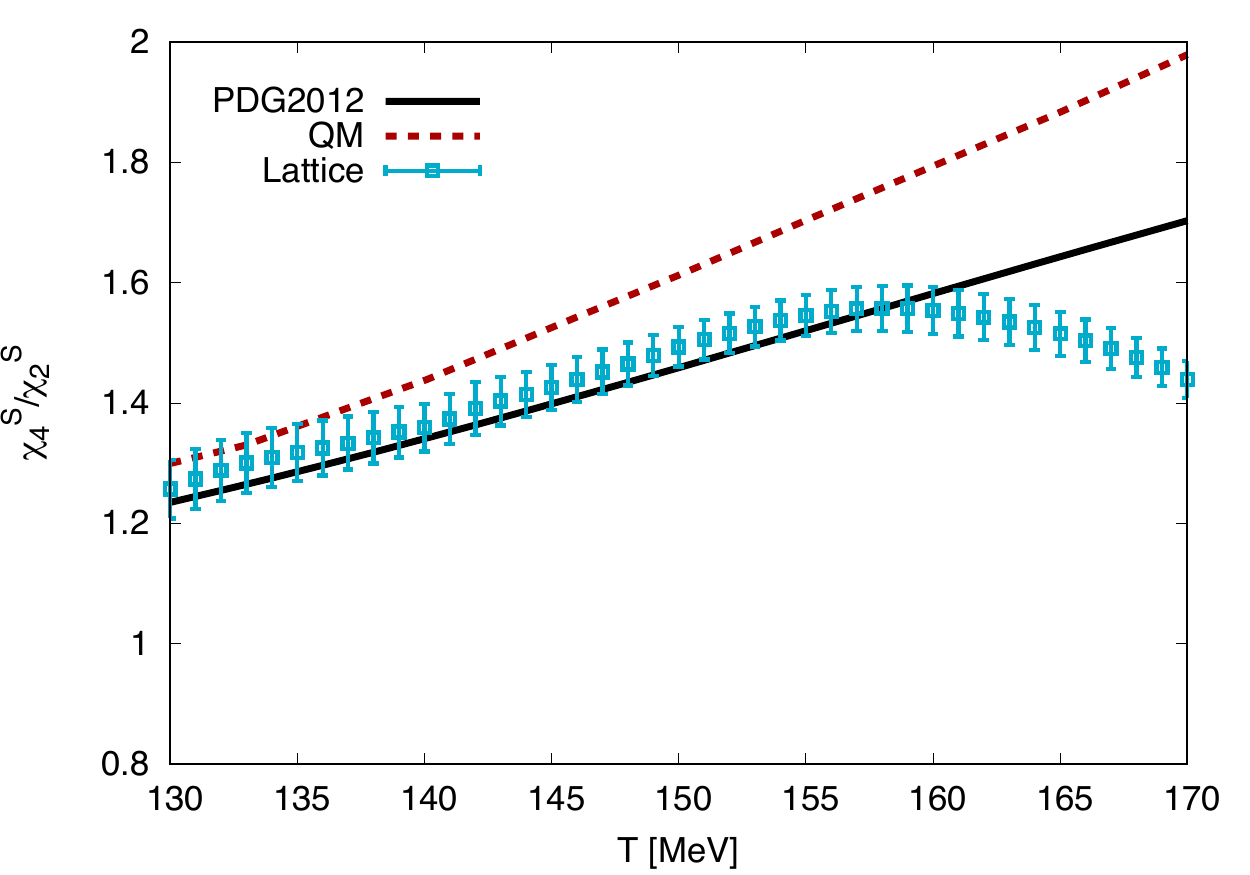}\includegraphics[width=14pc]{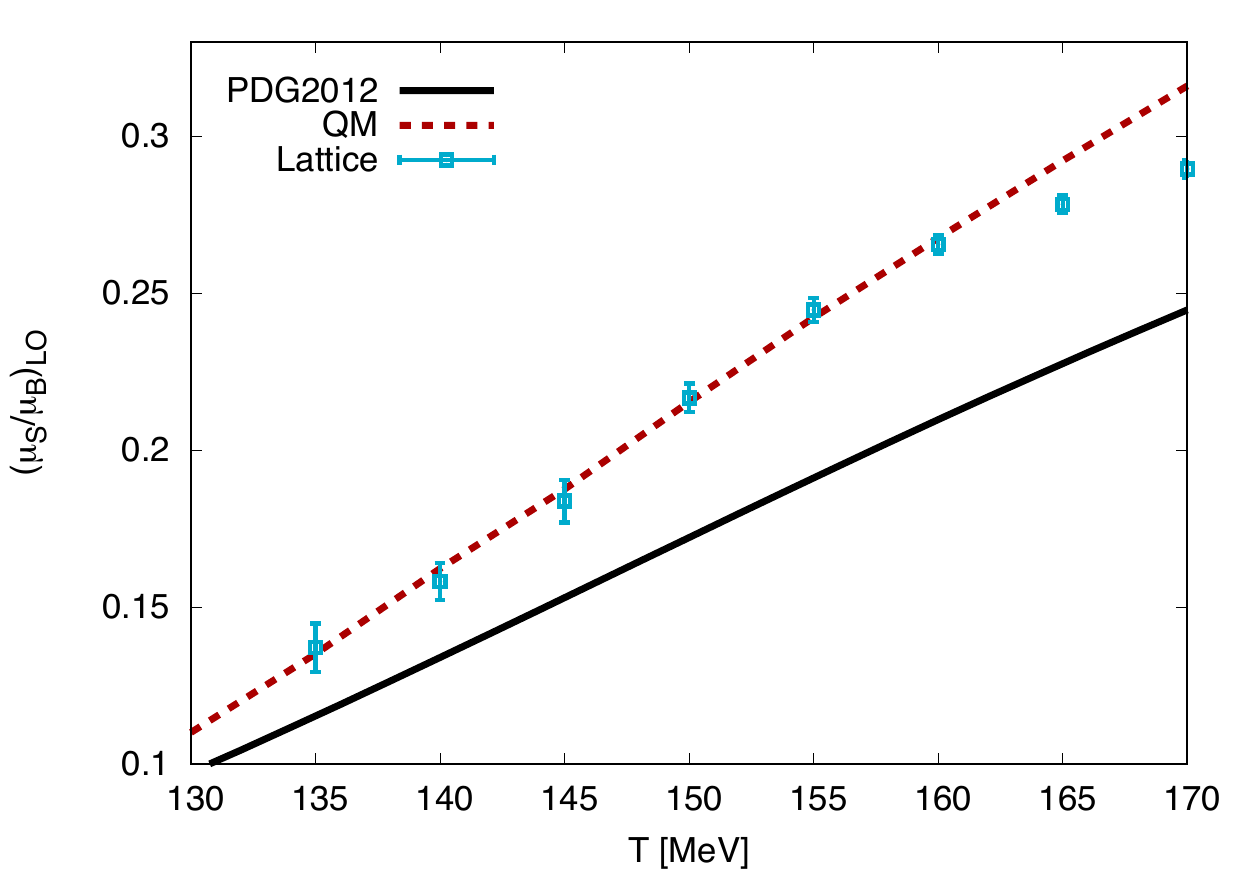}
\caption{\label{obs_old} Comparison of Lattice QCD and HRG model calculations based on the PDG2012 (solid black line) and QM (dashed red line) lists, for the ratio $\mu_S/\mu_B$ at leading order (left) and for the ratio $\chi^S_4/\chi^S_2$ (right).}
\end{figure}

With the increasing precision achieved in Lattice QCD calculations of more differential observables, some discrepancy with the HRG results appeared \cite{Bazavov:2014xya}. It was suggested that the inclusion of additional states in the spectrum, not yet experimentally discovered, but predicted by relativistic quark models (QM) \cite{Capstick:1986bm,Ebert:2009ub,Ferretti:2015ada} or Lattice QCD spectroscopy, might be necessary and cure the problem. However, the inclusion of additional states in the spectrum not always improves the agreement between HRG predictions and Lattice results \cite{Alba:2017mqu}, as can be seen in Fig.\ref{obs_old}.

\begin{figure}[h]
\center
\includegraphics[width=24pc]{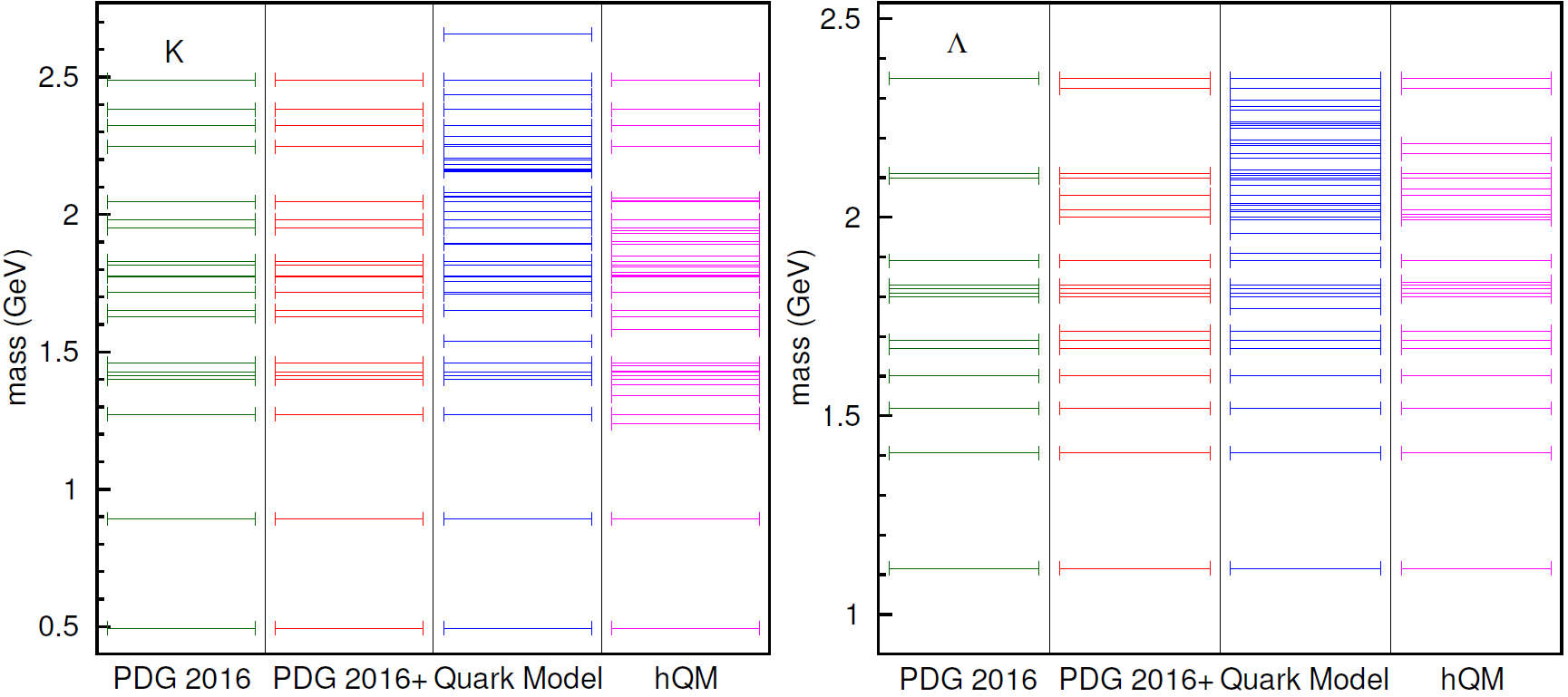}
\caption{\label{spectrum} Strange mesonic states and Lambda resonances experimentally established in the PDG2016 (green), PDG2016 including also one star states (red) and predicted by the QM (blue) and the hQM (magenta).}
\end{figure}

Therefore, a more systematic analysis of the content of the hadronic spectrum is needed, taking into consideration different possible choices. In Fig.\ref{spectrum} we compare, for illustrative reasons, the strange meson and Lambda states content of different lists: PDG2016 \cite{Patrignani:2016xqp} (all PDG states with **, *** and ****), PDG2016+ \cite{Patrignani:2016xqp} (including also * states), the original Quark Model (QM) \cite{Capstick:1986bm,Ebert:2009ub} and a more recent and refined hypercentral Quark Model \cite{Ferretti:2015ada} (hQM), where an interaction is included between the constituent quarks in the baryons. For hadrons containing only \textit{u}, \textit{d} and \textit{s} quarks, including both particles and anti-particles, as well as isospin multiplicity, the number of states for the lists extracted from the PDG are 608 in the case of PDG2016 and 738 in the case of PDG2016+ respectively. A large increase in the number of states comes from including QM resonances: the QM list contains 1517 states, whereas the hQM one contains 985. 

In the following, comparison between HRG and Lattice QCD results for several strangeness related observables is performed \cite{Alba:2017mqu}, which will depict a rather clear picture of whether, how many, and which states are most likely still missing in the hadronic spectrum established from experiment.

\section{HRG and Lattice: Partial pressures}

In the formulation of the HRG with no interactions, thermodynamical quantities can be calculated in the grand-canonical ensemble as a function of temperature and baryonic chemical potential \cite{Bluhm:2014wha}. The expression for the total pressure of the gas is
\begin{align*}
P_{tot}(T,{\mu}) &= \sum _k (-1)^{B_k+1} \dfrac{d_k T}{(2\pi) ^3} \int d^3\vec{p} \ln  \left(1+ (-1)^{B_k+1} \exp \left[ -\dfrac{(\sqrt{\vec{p} ^2+m_k^2}-\mu_k)}{T} \right] \right) \, ,
\end{align*}
where the sum runs over all hadrons and resonances included in the spectrum. Every quantity carrying the index $k$ depends on the specific state, hence on the specific term in the sum. The single particle chemical potential is defined with respect to the conserved charges of strong interactions (baryon number $B$, strangeness $S$ and electric charge $Q$) as $\mu_k = B_k \mu_B + S_k \mu_S + Q_k \mu_Q$. \\

In order to systematically analyze the different hadron spectra content, we separate the contribution to the pressures from hadrons grouped according to their quantum numbers. Under the assumption that Boltzmann approximation is valid for QCD thermodynamics in the range of temperatures below the QCD transition, it is possible to write the pressure of the system as
\begin{align*}
P(T, \hat{\mu}_B, \hat{\mu}_S) &= P^{BS}_{00}(T) + P^{BS}_{10}(T)\cosh(\hat{\mu}_B) + P^{BS}_{01}(T) \cosh(\hat{\mu}_S) + P^{BS}_{11}(T) \cosh(\hat{\mu}_B-\hat{\mu}_S) + \\
& \quad + P^{BS}_{12}(T) \cosh(\hat{\mu}_B-2\hat{\mu}_S) + P^{BS}_{13}(T) \cosh(\hat{\mu}_B-3\hat{\mu}_S)  \, ,
\end{align*}
where $\hat{\mu}_i = \mu_i/T$ and in $P^{BS}_{ij}(T)$, $i$ and $j$ refer to the absolute values of baryon number and strangeness respectively. The partial pressures can be calculated on the Lattice using an imaginary chemical potential approach. Indeed, setting $\mu_B=0$ and $\mu_S= i \mu_I$, the previous expression for the pressure becomes a sum of cosine terms, with the partial pressures being the Fourier coefficients. A combined fit of different observables yields the partial pressures as a function of the temperature. On the other hand, calculating the partial pressures with the HRG model is straightforward: it will be sufficient to sum over particles with the corresponding quantum numbers only. 

\begin{figure}[t]
\center
\includegraphics[width=15pc]{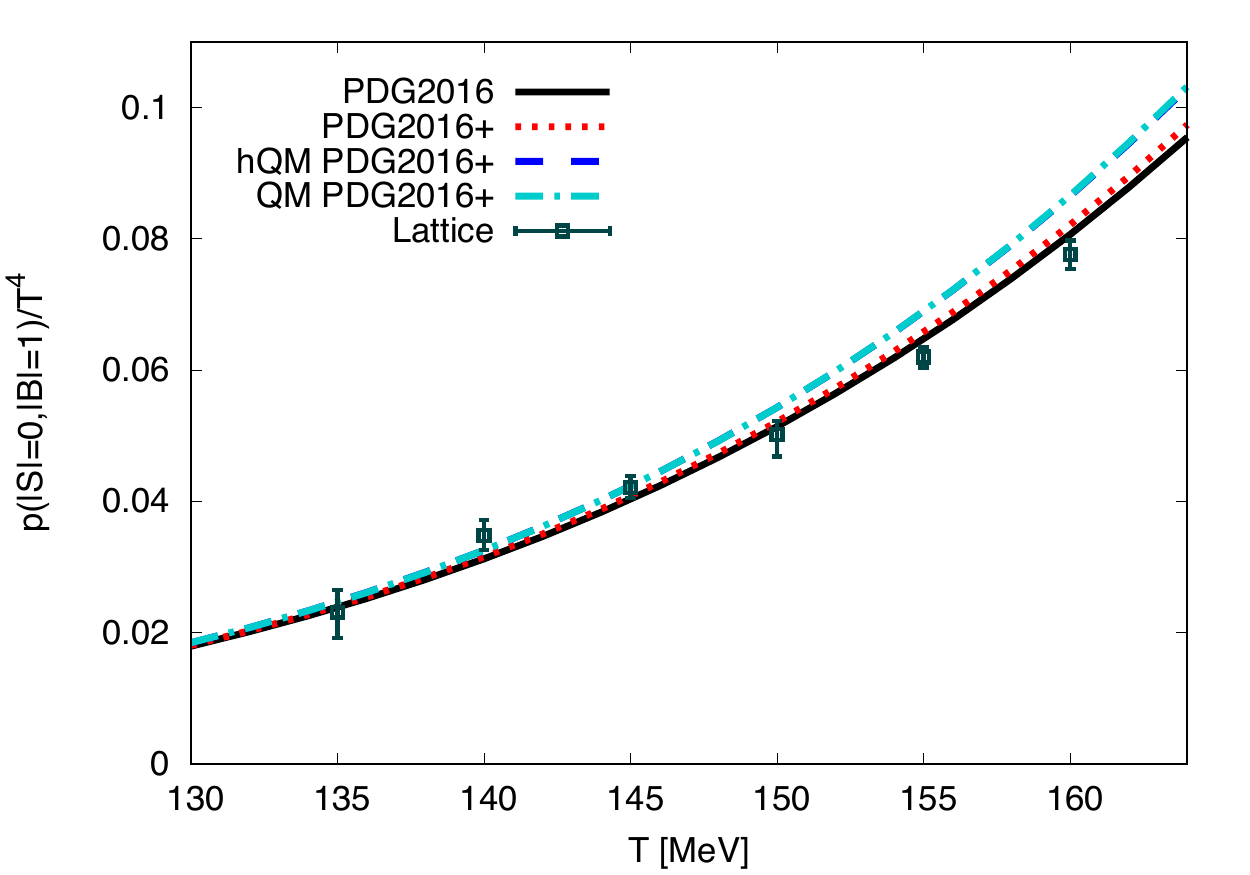}
\includegraphics[width=15pc]{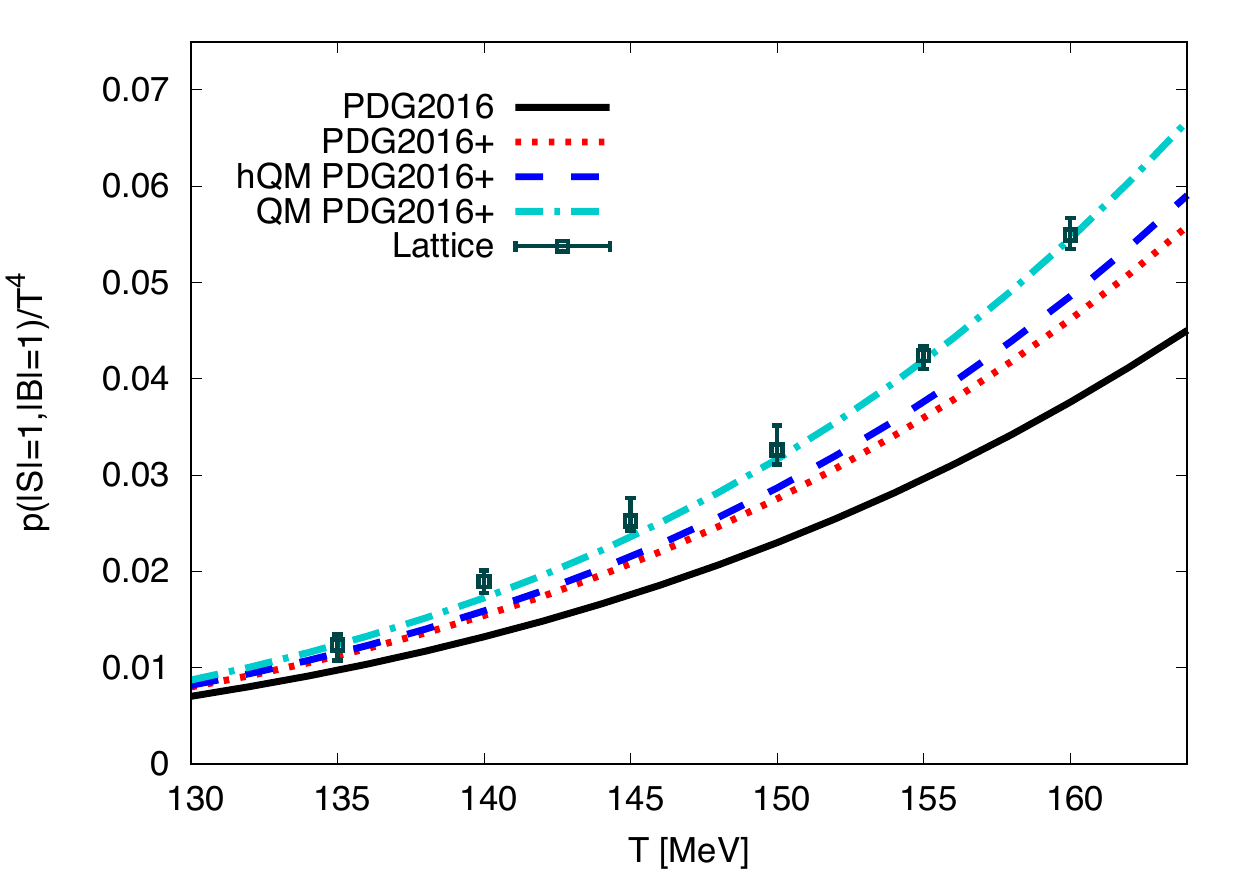} \\
\includegraphics[width=15pc]{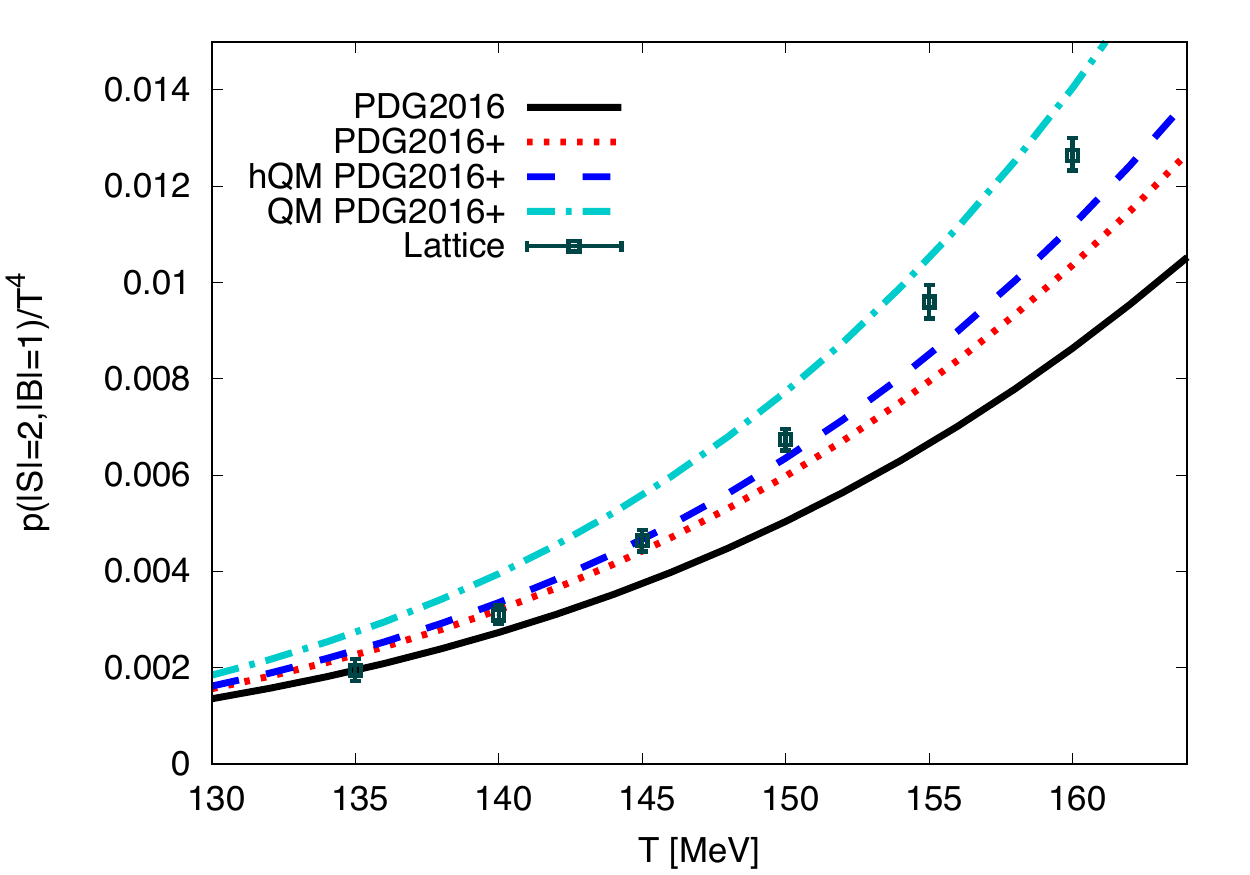}
\includegraphics[width=15pc]{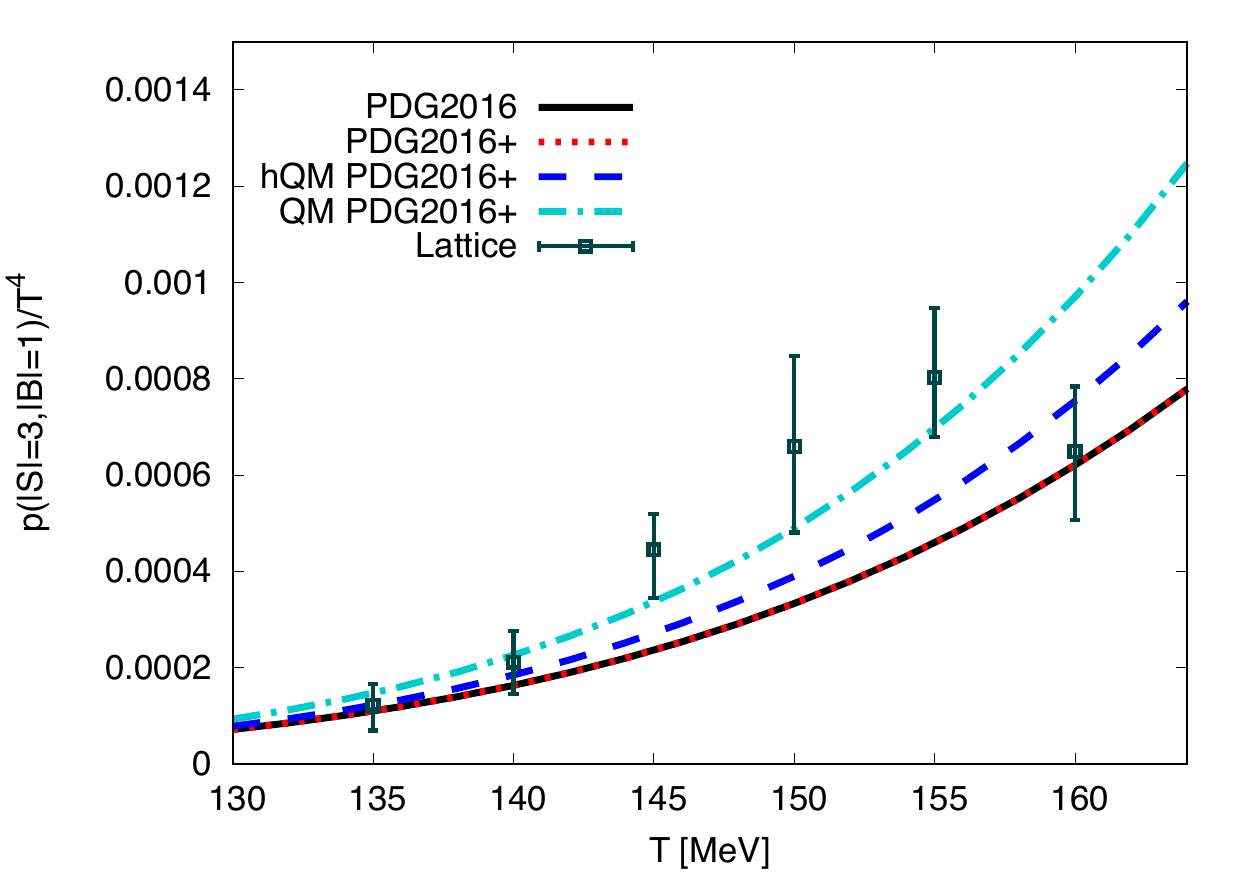}
\caption{\label{press_bar} Comparison of Lattice QCD results and HRG model calculations based on the PDG2016 (solid black line), PDG2016+ (dashed red line), hQM (dashed blue line) and QM (dot dashed light blue line) lists, for the partial pressures of $\left| S \right|=0$ (top left), $\left| S \right|=1$ (top right), $\left| S \right|=2$ (bottom left) and $\left| S \right|=3$ (bottom right) baryons. All Lattice results are continuum extrapolated.}
\end{figure}

\begin{figure}[h]
\center
\includegraphics[width=16pc]{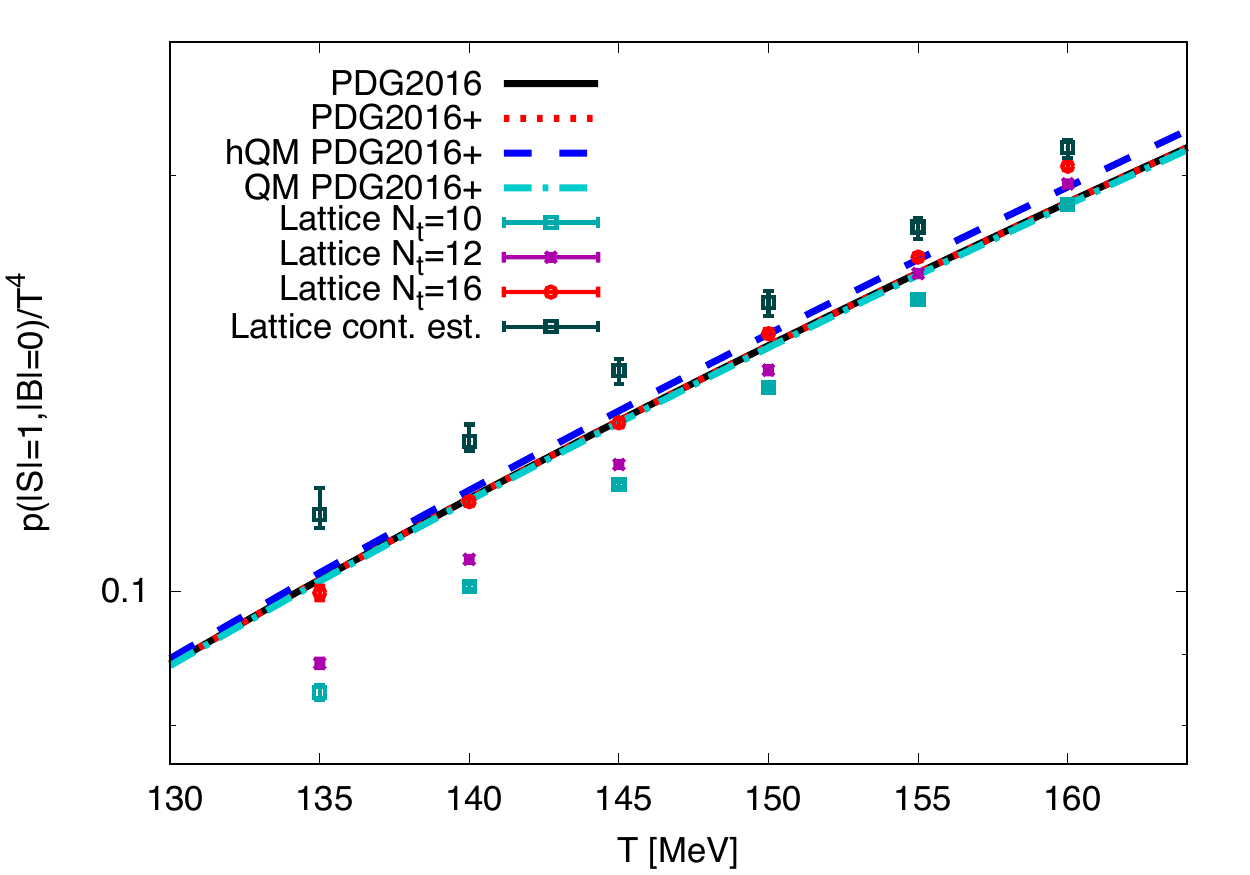}
\caption{\label{press_mes} Comparison of Lattice QCD results and HRG model calculations based on the PDG2016 (solid black line), PDG2016+ (dashed red line), hQM (dashed blue line) and QM (dot dashed light blue line) lists, for the partial pressures of $\left| S \right|=1$ mesons. The black dots indicate the continuum estimate.}
\end{figure}

\section{Results and discussion}
We show the comparison of Lattice QCD results and HRG model predictions using PDG2016, PDG2016+, QM and hQM lists, for the partial pressures of $\left| S \right|=0$,$\left| S \right|=1$,$\left| S \right|=2$ and $\left| S \right|=3$ baryons, as well as $\left| S \right|=1$ mesons. Note that, the pressure being a positive definite quantity, a prediction from the HRG model that underestimates the Lattice data in a certain sector, for a specific hadron list, would directly imply that the list does not contain enough states in the specific sector considered. Also, if the prediction overestimates the data, then it will follow that too many states are included in such a list.

In Fig.\ref{press_bar} we can see the partial pressure for different baryonic sectors. The first thing that catches the eye is that, except for the case on non-strange baryons, the most established states from the PDG are not enough to describe the data. For $\left| S \right|=1$ and $\left| S \right|=2$ baryons, the inclusion of * states sensibly improves the agreement with the data. When futher states from the hQM list are included, an additional slight improvement is noticeable in all the cases. For $\left| S \right|=1$ and $\left| S \right|=3$ baryons, however, it still seems that states predicted by the QM are needed to reproduce the data; on the other hand, these are clearly too many in the case of $\left| S \right|=2$ baryons.

Different considerations need to be made about the strange mesons in Fig.\ref{press_mes}. First, data are shown for different lattice spacings, and a continuous estimate is shown in place of a continuum extrapolation, which was not achievable being the data not in a clear scaling regime. Secondly, it is remarkable that all the lists seem to underpredict the value of this partial pressure. This may be due to two different reasons: either all the lists do not contain enough states in this sector, or the assumption of the ideal HRG model may not be applicable in the case of strange mesons.

\section{Conclusions}
In this proceedings we showed the analysis of different hadronic spectra through the comparison of Lattice QCD results and HRG model predictions. In the baryonic sector, it was rather evident that the list of established states listed by the PDG is not sufficient, and additional states are needed for all the families, in a couple cases even up to those from the original quark model. In the mesonic sector, it seemed that additional states are needed in all the considered spectra, if one trusts that the description of an ideal HRG model holds in the case of strange mesons too.

Through the presented analysis, it was possible to discern to what extent, and in what quantum number sectors, additional states need to be included in the hadronic spectra to those listed as established by the Particle Data Group. This can impact any analysis of the QCD transition in heavy-ion collisions that makes use of thermal models, such as the extraction of freeze-out parameters through thermal fits \cite{Andronic:2011yq,Cleymans:2005xv}, or ratios of fluctuations of conserved charges \cite{Borsanyi:2011sw,Bazavov:2012jq,Bluhm:2014wha}, especially in the study of strangeness freeze-out \cite{Bellwied:2013cta,Noronha-Hostler:2016rpd}.

\section*{Acknowledgements}
This material is based upon work supported by the National Science Foundation under Grants No. PHY-1513864, PHY-1654219 and OAC-1531814 and by the U.S. Department of Energy, Office of Science, Office of Nuclear Physics, within the framework of the Beam Energy Scan Theory (BEST) Topical Collaboration. An award of computer time was provided by the INCITE program. This research used resources of the Argonne Leadership Computing Facility, which is a DOE Office of Science User Facility supported under Contract No. DE-AC02-06CH11357. The authors gratefully acknowledge the use of the Maxwell Cluster and the advanced support from the Center of Advanced Computing and Data Systems at the University of Houston.

\section*{References}
\bibliography{FAIRness2017}

\end{document}